\documentclass[a4paper]{jpconf}
\usepackage{graphicx}
\usepackage[]{caption}
\usepackage{multicol}

  \usepackage{float,upgreek}
  \usepackage{subfigure}
\usepackage{booktabs}
\begin{document}
\title{Study of b-jet tagging performance in ALICE }

\author{Linus Feldkamp, for the ALICE Collaboration}

\address{Westf\"alische Wilhelms-Universit\"at M\"unster, Institut f\"ur Kernphysik, Wilhelm-Klemm Stra\ss e 9, 48149 M\"unster, Germany}

\ead{l.feldkamp@cern.ch}

\begin{abstract}
We present the current status and Monte Carlo study based performance estimates of b-jet tagging using ALICE, as obtained using both impact parameter as well as secondary vertex methods. We also address the prospects of the identification of electrons from
heavy-flavour hadron decays to obtain b-jet enhanced jet samples. 
\end{abstract}

\section{Introduction}
Charm and beauty quarks, produced in the early stage of heavy-ion collisions, are ideal probes to study the characteristics of the hot and dense deconfined medium (Quark-Gluon Plasma) formed in these collisions. 
The radiative energy loss of high energy partons interacting with the medium is expected \cite{[-1]}\cite{[0]} to be larger for gluons than for quarks, and to depend on the quark mass, with beauty quarks losing less energy than charm quarks, light quarks and gluons.
Therefore a comparison of the modification in the momentum distribution or possibly in the jet shape of b-jets with that of light flavour or c-jets in Pb-Pb collisions  relative to pp collisions allows one to investigate the mass dependence of the energy loss.
It also allows one  to study the redistribution of the lost energy and possible modifications to b-quark fragmentation in the medium. 
We present a MC study of the b-jet tagging performance in ALICE for pp collisions at $\sqrt{s} = 7\mathrm{\;TeV}$ using 
two different algorithms, both exploiting the long lifetime and high mass of B mesons to discriminate b-jets from jets from lighter partons.
\subsection{Monte Carlo sample and jet reconstruction}
In this analysis two sets of Pythia 6 \cite{[1]} (+GEANT 3 \cite{[2]}) simulations of proton-proton collisions  at  $\sqrt{s}=7$ TeV were used: $1.5 \times 10^7$  events with a  c-$\bar{\mathrm{c}}$ or b-$\bar{\mathrm{b}}$  pair and  $8.0 \times 10^7$  minimum bias  events.
Jets were reconstructed using FastJet \cite{[3]} v2.4.2 with the anti-$k_\mathrm{T}$ algorithm \cite{[4]}, a cone radius of $R=0.4$ and a minimum track $p_T$ of $150$ MeV$/c$. The jet reconstruction was done both on reconstructed tracks and at generator level, matched by the minimal $\Delta R =\sqrt{(\eta_{Jet,Gen.}-\eta_{Jet,Rec.})^2+(\varphi_{Jet,Gen.}-\varphi_{Jet,Rec.})^2}$. At generator level neutral particles were considered to correct to the full jet energy. Jets with $p_{\mathrm{T}}>10$\;GeV$/c$ and a charged area larger than $0.6\pi R^2$ were taken into account for further analysis steps. The jet flavour is determined by the flavour of the leading $p_\mathrm{T}$ parton in the cone. In case it is a gluon, the daughters are also considered to take gluon splitting into account.

\subsection{Track counting algorithm}
 The \textit{track counting algorithm} \cite{[5]}  exploits the large $r\varphi$-impact parameters, $d_0 = \vert \vec{d}_0 \vert $, of B-meson decay products to identify  b-jets. The signed $r\varphi$-impact parameter, $d_0=sign(\vec{d}_0 \cdot \vec{p}^{Jet}) d_0$, is calculated for each track in the jet cone, where $\vec{d}_0$ is pointing away from the primary vertex. The tracks are required to have a $p_\mathrm{T}> 1\;$GeV$/c$, a hit in the innermost layer of the ITS and a Distance of Closest Approach between track  and jet smaller than 0.07 cm. To tag a jet as a b-jet, the $d_0$ parameter of each track in the jet is ordered by size, and the N-th largest $d_0$ is compared to check if it is larger than a threshold value $a$. In the analysis presented here N=3 and the threshold value a was varied in such was to keep the same b-tagging efficiency for different jet $p_\mathrm{T}$ bins.
 The tagging efficiency for a jet assigned to a given parton, c, b, LFg=(u,d,s,g), can be defined as: 
 \begin{equation} 
  \epsilon_i(\mathrm{p}^{Jet}_\mathrm{T}) = \frac{N^{tagged}_i (\mathrm{p}^{\mathrm{Jet}}_\mathrm{T}) }{N_i (\mathrm{p}^{\mathrm{Jet}}_\mathrm{T}) } \quad \mbox{ with $i$ =c,b, LFg(=u,d,s,g)}
 \end{equation}
Tagging efficiencies for c- and b-jets and the ratio $\epsilon_c/\epsilon_b$ (at constant $\epsilon_b$) for the N=3 discriminator were obtained as a function of the momentum of the jet at kinematic level, thus including the neutral component. 
To allow a comparison of the efficiencies of charm-, beauty- and LFg-Jets the ratio 
$\frac{\epsilon_i }{\epsilon_b}\vert_{\epsilon_b(\mathrm{p}_\mathrm{T})\mbox{\small \textnormal \textit{fixed}}}$ is used.
In Figure 1 we show the results for the case $\epsilon_b(\mathrm{p}_\mathrm{T})=0.1$, which corresponds to  threshold $a$ on the 3rd most displaced track of about 0.01 cm. It may be noted that $a$ is varied with $p_\mathrm{T}$ in order to achieve a constant $\epsilon_b$.
\begin{figure}[H]
\centering
 \begin{subfigure}[] 
  {
     \includegraphics[width=7.5cm]{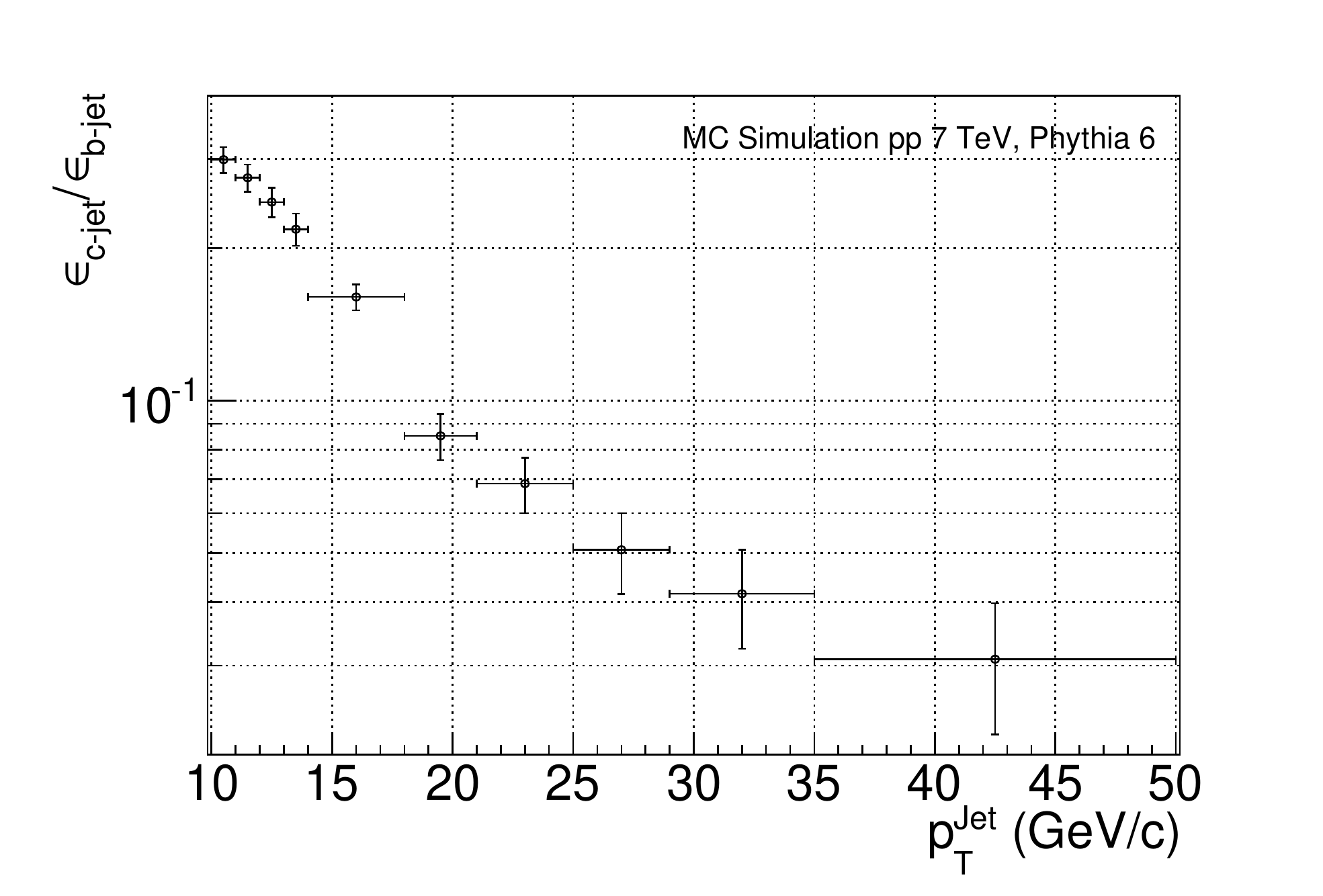} 

  }
 \end{subfigure} 
 \begin{subfigure}[ ]
  {
    \includegraphics[width=7.5cm]{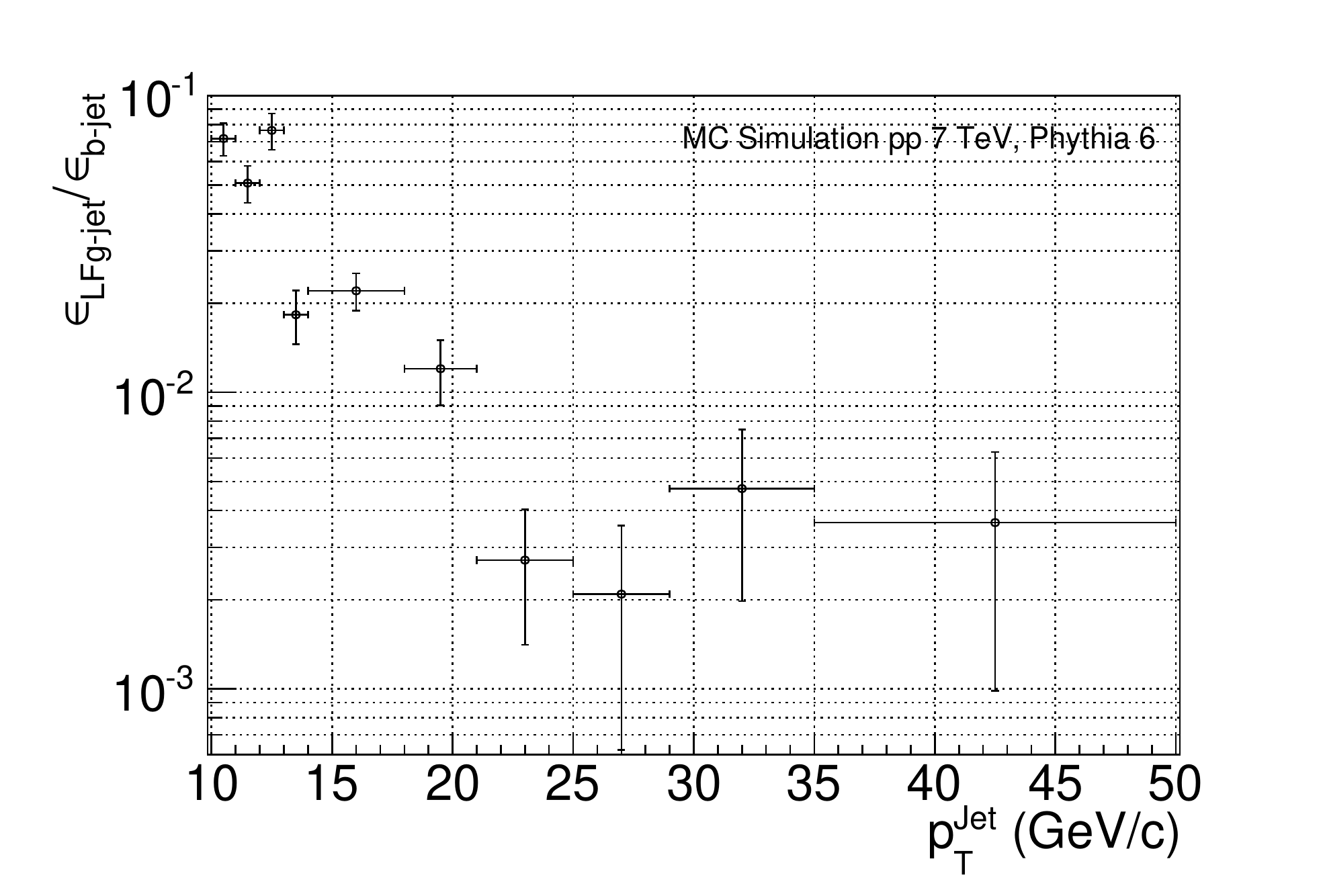} 

 }
 \end{subfigure}
  \caption{\footnotesize c-jet vs. b-jet performance for N=3, \textbf{(a)}: $\epsilon_{c}(\mathrm{p}_\mathrm{T})$/$\epsilon_b(\mathrm{p}_\mathrm{T})$ , \textbf{(b)}: $\epsilon_{LFg}(\mathrm{p}_\mathrm{T})$/$\epsilon_b(\mathrm{p}_\mathrm{T})$  at  $\epsilon_b(\mathrm{p}_\mathrm{T})$ = $0.1$.}

\end{figure}
\noindent
At $\epsilon_b=0.1$ the efficiency ratio $\epsilon_c/\epsilon_b$, obtained from the b- and c-quark enhanced Monte Carlo sample, is decreasing between 20 GeV/c  and 50 GeV/c from $\sim 0.08$  to $\sim 0.03$ for the N=3 discriminator. \\
%
\noindent
The light-flavour and gluon jets have typically tracks with smaller impact parameters with respect to charm jets. Thus the $\epsilon_{LFg}/\epsilon_b$ efficiency ratio, obtained from the minimum bias sample, is approximately one order of magnitude smaller, below $0.005$ above 20 GeV/$c$.
\subsection{Secondary vertex reconstruction}
In the \textit{secondary vertex algorithm} \cite{[6]}   secondary vertices are reconstructed combining 2 or 3 jet tracks, approximating the tracks as straight lines in the vicinity of the primary vertex. The presence of vertices with a displaced topology and with a large invariant mass is used as a criterion to tag jets from b. In particular, the discriminator variables used are the vertex invariant mass and the decay
length in the transverse plane, $L_{xy} = L_{xy} \cdot sign(\vec{L}_{xy} \cdot \vec{\mathrm{p}}_{\mathrm{T,Jet}})$.

\begin{minipage}[c]{1.\textwidth}
 
\begin{figure}[H]
\centering
\begin{subfigure}[ ]
{
  \includegraphics[height=5.0cm,width=4.95cm]{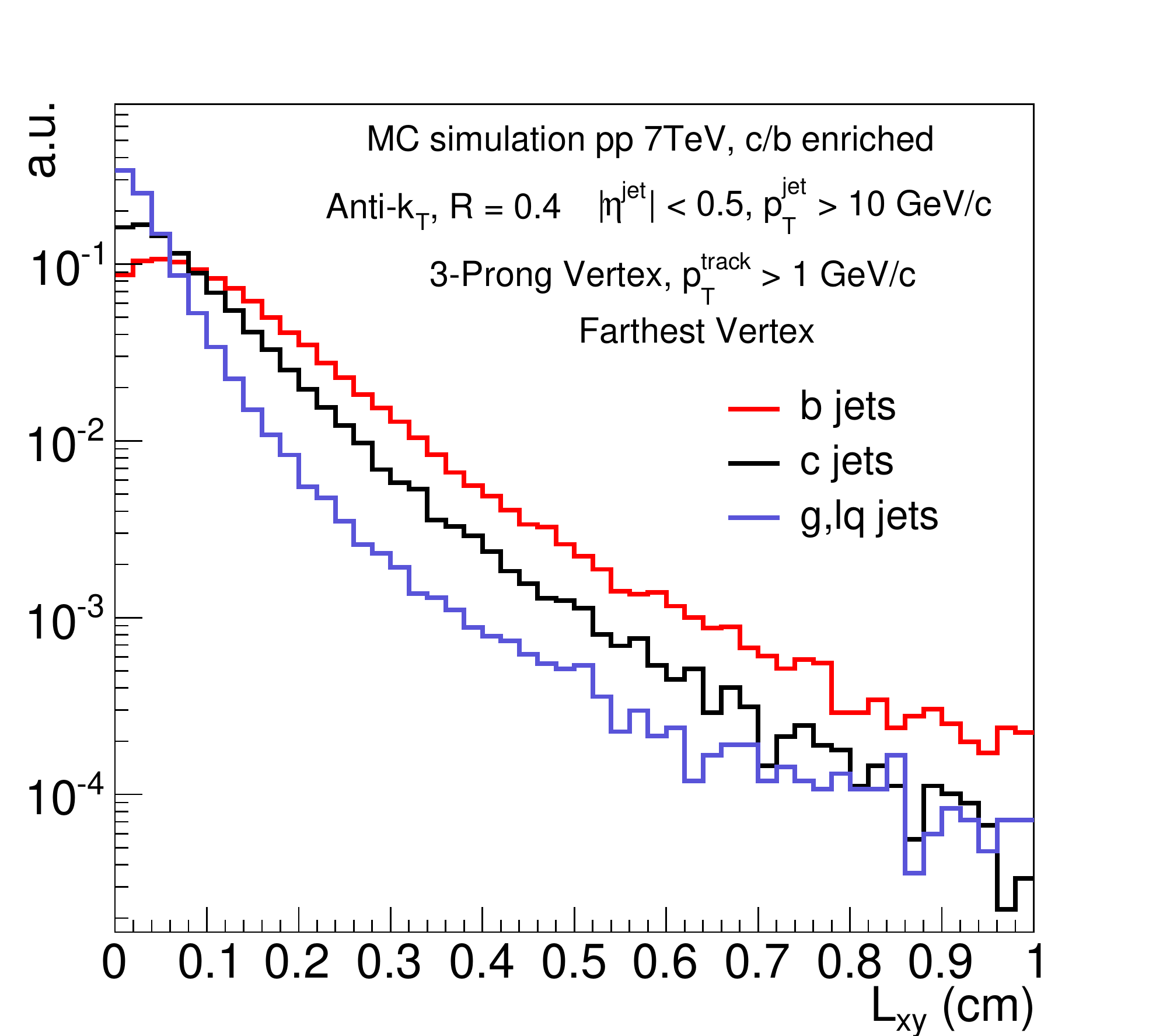}  }
 \end{subfigure}
 \begin{subfigure}[]
{
  \includegraphics[height=5.0cm,width=4.95cm]{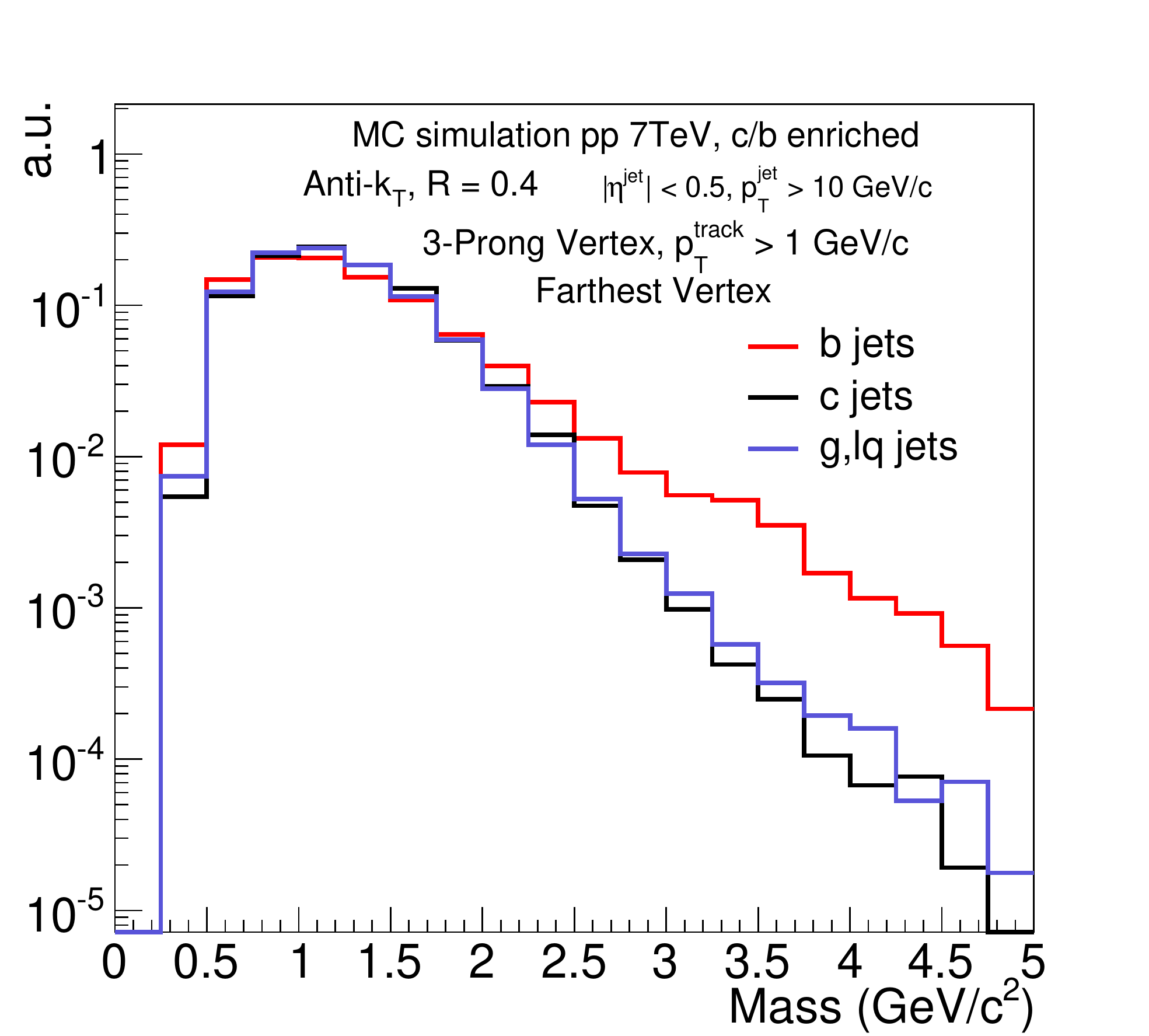}  
   }
 \end{subfigure}
 \begin{subfigure}[]
{ \includegraphics[height=5.0cm,width=4.95cm]{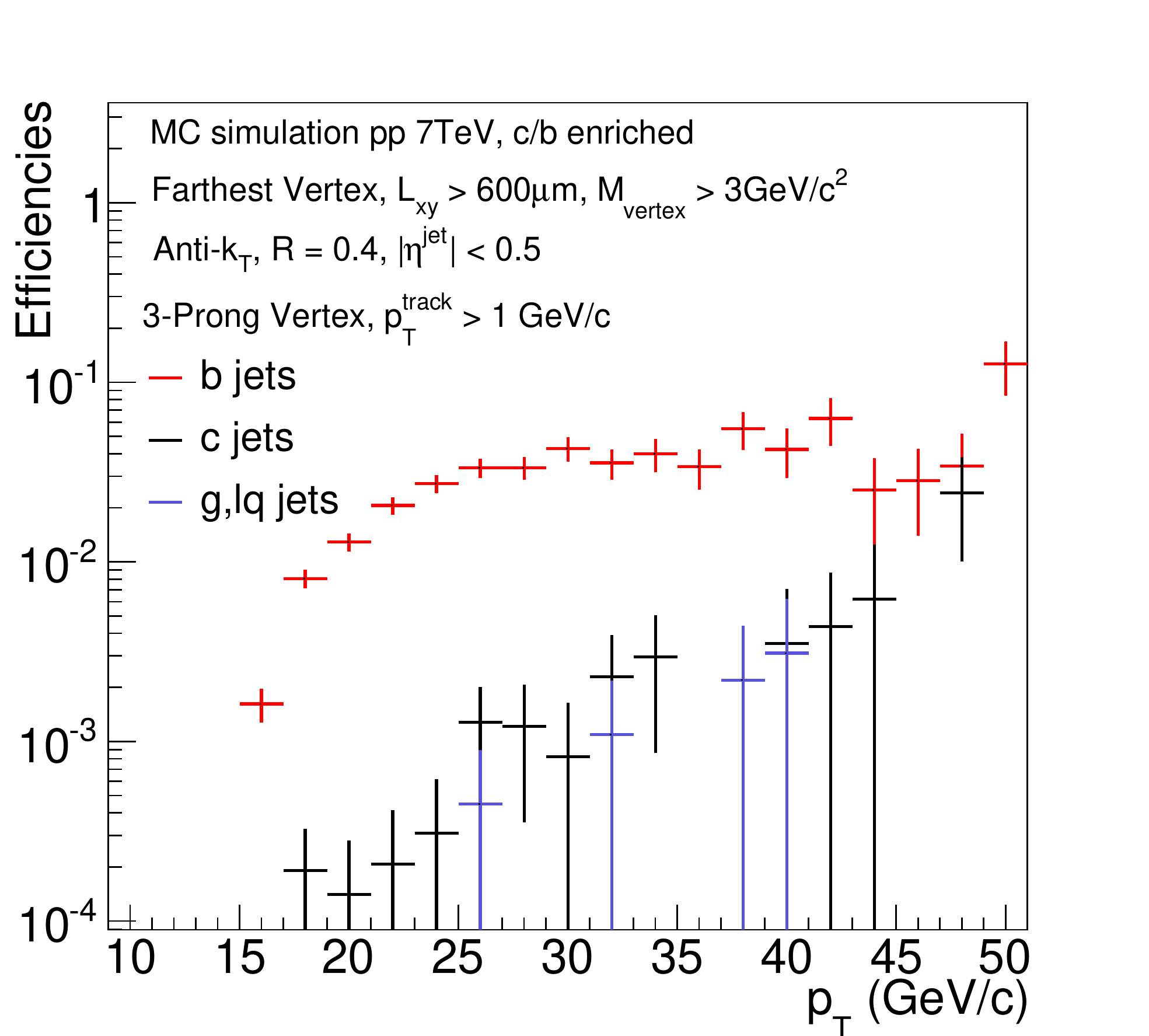}}
 \end{subfigure}
    \label{fig:fig3}
   \caption{\footnotesize\textbf{(a)}: $L_{xy} = L_{xy} \cdot sign(\vec{L}_{xy} \cdot \vec{\mathrm{p}}_{\mathrm{T,Jet}})$ using the most displaced vertex in the jet, \textbf{(b)}: Vertex invariant mass for the most displaced vertex in the jet, \textbf{(c)}: Tagging\;efficiencies: $ {{L_{xy}>600}\;\upmu\mbox{m, }} {M_{\mbox{Vertex}}>3}\;\mbox{GeV/}c^2$}

\end{figure}
\end{minipage}
\noindent    \\
\\
$L_{xy}$ and the vertex invariant mass distributions are shown in Figure 2a and 2b. In Figure 2c the tagging efficiency for a given combination of cut values is shown exemplarily. 

\subsection{Electron Identification}
The performance of b-jet tagging may further improve by using electron triggered events, as semi-electronic  decays of heavy-flavour hadrons have branching ratios of the order of 10\% \cite{[7]}.
The inclusive electron spectrum contains a significant contribution from heavy-flavour hadron decays above momenta of a few GeV/$c$
with the beauty-decay contribution becoming dominant above $\sim 5\;\mathrm{GeV/}c$ \cite{[8]}.
Thus using both the TPC and the electromagnetic calorimeter (EMCal) for electron identification
in the analysis and the high-$p_\mathrm{T}$ EMCal triggers a significant increase in b-jet statistics may be achieved by applying a 
 TPC-d$E$/d$x$ cut of $]-1\sigma,3\sigma[$ around the expected electron d$E$/d$x$, with $\sigma$ corresponding to the gaussian d$E$/d$x$-resolution and requiring a EMCal cluster energy $E$ over track momentum $p$ ratio of $0.8 < E/p < 1.2$.

\section{Conclusions}
Flavour-dependent tagging efficiencies for the track counting and the secondary vertex algorithm have been estimated using simulated pp events at  $\sqrt{s} = 7\mathrm{\;TeV}$. Beside further optimisation these efficiencies can be used as  input to estimate the contamination of the raw tagged sample of b-jets using FONLL \cite{[9]} or POWHEG \cite{[10]}. 
High-$p_T$ electron triggers promise a significant increase in statistics. A measurement of the b-jet spectrum in pp and p-Pb collisions down to $\approx 20$ GeV/$c$  and of the b-jet spectrum in Pb-Pb collisions down to $30-40$ GeV/$c$ might become feasible in the future.

\section*{References}
\begin{multicols}{2}

                     \end{multicols}

\end{document}